\documentclass[ams,twocolumn, amssymb,floatfix,nofootinbib]{revtex4-1}

\usepackage{amsmath}
\usepackage{tabularx}
\usepackage{bm}
\usepackage{slashed}
\usepackage{euscript}
\usepackage{graphicx}
\usepackage{color}
\usepackage{amsfonts}
\usepackage{exscale}
\usepackage{amsbsy}
\usepackage{subfigure}
\usepackage{textcomp}
\usepackage{comment}
\usepackage{hyperref}
\newcommand{\beq}{\begin{equation}}
\newcommand{\eeq}{\end{equation}}
\newcommand{\be}{\begin{eqnarray}}
\newcommand{\ee}{\end{eqnarray}}

\begin{document}

\title{Theory of superconductivity in doped quantum paraelectrics}
\author{Yue Yu$^{1,2}$, Harold Y. Hwang$^{1,2}$, S. Raghu$^{1,2,\ddagger}$, Suk Bum Chung$^{3,4,\dagger}$}
\affiliation{$^1$Stanford Institute for Materials and Energy Sciences,
SLAC National Accelerator Laboratory, Menlo Park, California 94025, USA}
\affiliation{$^2$Geballe Laboratory for Advanced Materials,
Stanford University, Stanford, California 94305, USA}
\affiliation{$^3$Department of Physics and Natural Science Research Institute, University of Seoul, Seoul 02504, Republic of Korea}
\affiliation{$^4$School of Physics, Korea Institute for Advanced Study, Seoul 02455, Republic of Korea}
\email{$^\ddagger$sraghu@stanford.edu}
\email{$^\dagger$sbchung0@uos.ac.kr}
\date{\today}

\begin{abstract}
Recent experiments on Nb-doped SrTiO$_3$ have shown that the superconducting energy gap to the transition temperature ratio maintains the Bardeen-Cooper-Schrieffer (BCS) value throughout its superconducting dome. Motivated by these and related studies, we show that the Cooper pairing mediated by a single soft transverse-optical phonon is the %only viable 
most natural mechanism for such a superconducting dome given experimental constraints, and present the microscopic theory for this pairing mechanism.  Furthermore, we show that this mechanism is consistent with the $T^2$ resistivity in the normal state. Lastly, we discuss what physical insights SrTiO$_3$ provides for superconductivity in other quantum paraelectrics such as KTaO$_3$.   
\end{abstract}

\maketitle

\section{ Introduction } The observation of superconductivity in `quantum paraelectrics' - materials with low temperature incipient ferroelectricity  -  has raised several fundamental questions on the pairing mechanism in such systems.  Examples of quantum paraelectrics include strontium titanate (SrTiO$_3$) \cite{Schooley1964,Koonce1967,XLin2014,Swartz2018,Thiemann2018,Yoon2021} %[WE SHOULD ALSO CITE Phys. Rev. Lett. 120, 237002 here and below - microwave conductivity experiments showing the gap to Tc ratio.], 
as well as potassium tantalate (KTaO$_3$) \cite{Ueno2011,ZChen2021L,CJLiu2021,ZChen2021S} and lead telluride (PbTe)\cite{Matsushita2006}. A central question involves the hierarchy of the relevant fluctuations and their consequences for superconductivity.  For instance, if pairing is mediated mainly by soft critical ferroelectric fluctuations, the associated superconducting dome would be confined to low electron densities, as ferroelectricity itself is sharply defined only in an insulating phase (A `ferroelectric metal' is essentially %one with 
characterized by either broken inversion, {\it i.e.} non-centrosymmetric, or spatial reflection symmetries \cite{Anderson1965}; dipole moments while permitted by symmetry, are strongly screened by the conduction electrons.).   The fact that superconductivity, at least in niobium-doped strontium titanate, is observed only over a range in the  dilute limit ($\sim$0.05\% to $\sim$0.5\%)\cite{XLin2014,Swartz2018,Yoon2021} gives support to the notion of pairing mediated by ferroelectric fluctuations \cite{Enderlein2020}.  

%HYH: I think I understand why you say 'astonishing' in a normal el-ph SC sense, but for me this sounds strong given the dome in STO spans at least ~1 order of magnitude, compared to domes in the cuprates, pnictides, etc. 

The restriction %to 
of pairing %in a narrow range of 
to dilute carrier concentrations, however, presents several puzzling issues.  First, the resulting %vanishing 
small density of states in a %dilute 
3{\it d} electron system would suggest a correspondingly small superconducting pairing strength. Second, the soft mode associated with ferroelectricity is the transverse optical (TO) phonon, which couples %quite weakly 
%cannot have a strong conventional coupling to the electrons. %; for instance, 
less strongly to the electrons than the longitudinal optical (LO) phonons. 
%Indeed, 
Moreover, symmetry considerations lead to the conclusion that in the absence of orbital- or spin-dependent processes, electrons can only scatter off of pairs of TO phonons \cite{vanDerMarel2019,Gastiasoro2020A,Volkov2021}.  The resulting reduction in phase space would naturally result in reduction of the superconducting transition temperature $T_c$. %(SBC: Please check - we want Sec III second, third, fourth paragraphs to sound like we are revisiting these sentences) {\bf do we actually mean ``in the absence of orbital-dependent process?"--YY} 
Finally, given the dilute electron concentrations, there is the possibility that the Fermi energy may be smaller than the phonon frequency itself, resulting in an inverted `anti-adiabatic' pairing regime.  Whether superconducting domes can arise in quantum paraelectrics despite these circumstances remains actively debated \cite{Migdal1958,Eliashberg1960,Pietronero1995,Grimaldi1995,Chubukov2020}. %(SBC: any additional citation?).  
%The origin of the superconducting dome, the effect of soft ferroelectric fluctuations, in conjunction with the limited electronic density of states available for superconductivity, has led to several fascinating conjectures for  pairing mechanisms. 
Moreover, the observation of superconductivity at interfaces of quantum paraelectrics \cite{Reyren2007,Valentinis2017,Ueno2011,ZChen2021L,CJLiu2021,ZChen2021S} further motivates the study of superconductivity in these systems, raising the question of the role of spatial dimensionality on all these issues.  

%Two popular scenarios for pairing mechanisms have emerged in light of the experimental observations.  First, as a consequence of the dilute limit, if electronic energy scales are comparable to, or lower than the Debye energies, superconductivity emerges in an inverted `anti-adiabatic' regime.  In this scenario, it is difficult to make the case for a soft mode, since by definition, a quantum critical fluctuation has an energy scale parametrically smaller than bare electronic scales.  Second, there is the  possibility that superconductivity is more conventional in nature, with a few low lying phonon modes being responsible for pairing, and whose energy scales are substantially smaller than the Fermi energy.  If the latter were true, the pairing mechanism would be identical in nature to a retarded attractive force mediated by the electron phonon coupling, and it remains a challenge to explain why a dome occurs in this scenario.  Which of these ideas is the most likely one for most quantum paraelectrics remains unclear and our goal here is to disentangle these possibilities.  

In this Letter, we construct a self-consistent pairing scenario for quantum paraelectrics, and illustrate it in the case of %strontium titanate, 
SrTiO$_3$, where recent experiments have placed significant constraints on theory.  %[DELETED A SENTENCE - IF YOU WANT TO INCLUDE, THERE ARE MANY EXPERIMENTS CONTRIBUTING - TUNNELLING, OPTICS, TRANSPORT, NEUTRONS, ETC. - THEN THESE SHOULD BE CITED - SEE HYEOK'S PAPER. NOTE THAT CONVENTIONAL TUNNELING ONLY "SEES" LO PHONONS - WE DO NOT SEE TO MODES - IN PRINCIPLE WE SHOULD SEE THESE BY YOUR MECHANISM, BUT IT IS SELF-CONSISTENT THAT WE DO NOT SEE THEM SINCE THEY ARE WEAK COUPLING.] 
These experiments have reported a textbook BCS gap to $T_c$ ratio \cite{Swartz2018,Thiemann2018,Yoon2021}; any theory of pairing in this system must satisfy this constraint.  In light of these experiments, we discuss constraints on pairing that arises from either the anti-adiabatic or the more conventional adiabatic pairing mechanisms.  We construct a scenario in which pairing is mediated by TO phonons.  We also present a mechanism by which electrons may couple to single TO phonons, resolving the issues of limited phase space alluded to above.  
We then discuss the relevance of these findings to other quantum paraelectrics, including interfacial systems.  
%  

%HYH: For STO I think we can make a stronger statement - only LO4 is always above EF for SC samples. At the overdoped boundary, EF is around 65 meV (while LO4 is 100 meV) - already this is certainly not 'greatly exceeded'. For LO3,LO2,LO1 (where LO3 and LO4 el-ph coupling is also clearly large), EF crosses them within the dome itself.

%By contrast, a conventional adiabatic pairing scenario can only arise from the lowest energy transverse optic (TO1) phonon, which is the soft mode associated with nearby ferroelectricity.  While we discuss both scenarios, we focus greater depth on the TO pairing mechanism, construct a microscopic multiband model to account for a superconducting dome.  We then comment on normal state electrical properties stemming from TO pairing mechanisms.  

\section{Results}
\subsection{Experimental considerations}

Two distinct pairing scenarios have been proposed to explain superconductivity in the dilute limit of bulk %strontium titanate: 
SrTiO$_3$: a conventional one in which the phonon frequency remains smaller than the Fermi energy $E_F$ \cite{Gurevich1962,Kedem2018,Kozii2019,Gastiasoro2020B}, and an anti-adiabatic mechanism in which the hierarchy of energy scales are inverted \cite{Appel1969,Takada1980,Edge2015,Gorkov2016,Ruhman2016,Woefle2018,Gastiasoro2019,Klimin2019}.  In recent experiments \cite{Swartz2018,Yoon2021}, various phonon frequencies were probed, in addition to the superconducting gap. Various experiments also show that 1) the lowest TO (TO1) phonon frequency increases with doping but remains below the Fermi energy across the superconducting dome \cite{Bauerle1980,vanMechelen2008,Yoon2021}, 2) the longitudinal optical (LO) phonon frequencies remain unchanged with doping and are either comparable to or greater than $E_F$ across the dome \cite{Choudhury2008}, and 3) the superconducting gap to $T_c$ ratio is close to the BCS value \cite{Swartz2018,Thiemann2018,Yoon2021}.  

It follows from the first observation that any pairing mechanism involving the TO1 phonons can remain conventional and adiabatic 
%(in contrast to the assumption made by Volkov {\it et al.} \cite{Volkov2021}) 
whereas LO pairing mechanisms would be in the anti-adiabatic regime in %strontium titanate.  
SrTiO$_3$. We first briefly describe why the anti-adiabatic scenarios are unlikely in %strontium titanate 
SrTiO$_3$ and we then consider the adiabatic pairing scenario mediated by TO phonons.

Based on the tunneling measurements \cite{Swartz2018,Thiemann2018,Yoon2021}, any anti-adiabatic pairing scenario that remains viable across the superconducting dome must necessarily only involve the highest LO phonon mode (LO4).  Furthermore, the constraint imposed by the BCS gap to $T_c$ ratio requires the effective attraction mediated by the LO phonon to be weak, which can occur in the anti-adiabatic regime only if  the LO4 phonon frequency were significantly higher than the Fermi energy.    Further restrictions from the tunneling data come from the fact that the LO frequency remains essentially unchanged with doping.  Since the BCS coupling is proportional both to the density of states and the square of electron-phonon coupling, a crucial ingredient needed for $T_c$ to decrease beyond  optimal doping is the reduction of the electron-phonon coupling strength with %doping 
the dopant concentration $n$ faster than $n^{-1/3}$, in order to overcome the growth of the BCS coupling with increasing density of states.  It therefore seems unlikely, based in part on the tunneling measurements, that pairing in %STO 
SrTi$_{1-x}$Nb$_x$O$_3$ is mediated by an anti-adiabatic LO4 phonon.  

%The above issues with the possible anti-adiabatic pairing mechanisms in 
We are thus led naturally to consider an adiabatic pairing mechanism across the dome of SrTi$_{1-x}$Nb$_x$O$_3$.  The only phonon mode that remains in the adiabatic regime across the dome is the TO1 mode, the softening of which leads to ferroelectricity.  Furthermore, a conventional BCS framework based on the Migdal approximation should suffice to account for the BCS gap to $T_c$ ratio within this scenario (The lower density dome that is observed in oxygen-reduced samples \cite{XLin2014} is outside the scope of the present paper as such samples have been resistant to the pairing gap measurement through the planar tunneling spectroscopy.).  

Additionally, the superconducting dome from TO1 phonon exchange can be simply understood as follows.   Prior experiments\cite{Bauerle1980,vanMechelen2008} indicate that the TO1 phonon frequency increases with carrier concentration as 
%.  demand an attempt to construct adiabatic pairing mechanism. For the doping range where the superconducting dome, the only adiabatic phonon, or any bosonic, mode left is the TO1 phonons, the softening of which leads to ferroelectricity. 
%The superconducting dome with the BCS ratio is a guaranteed feature of the adiabatic pairing mediated by TO1 phonons if one takes into account the experimental data %on the evolution of the Fermi energy and 
%which shows the square of the TO1 mode energy gap increasing linearly with the Nb doping, {\it i.e.} 
$\omega_T^2 = K_0 + n K_1$ with the approximate values of $K_0 \approx$ 1meV$^2$ and $K_1 \approx 1.8 \times 10^{-19}$meV$^2$cm$^3$ $> 0$. \cite{Gastiasoro2020A} %[SHOULD CITE THE ORIGINAL PAPERS - IT IS NOT OUR TUNNELING WORK - SEE HYEOK'S PAPER.] 
%Given that 
Hence 
%the single-band 3D Fermi liquid density of state $N_F$ should be proportional to %the cubic root of the carrier density, 
%$n^{1/3}$, 
the BCS eigenvalue for the adiabatic pairing mediated solely by a single TO1 phonon is parametrically
\begin{equation}
\lambda_{\rm BCS} \propto \frac{N_F}{\omega^2_T} \sim \frac{n^{1/3} }{K_0 + K_1 n};
\label{singleBand}
\end{equation}
the overdoped attenuation of $T_c$ naturally comes from the fact that the TO1 phonon hardens `faster' with Nb concentration than the increase in the density of states.  Thus, in the adiabatic pairing scenario based on TO1 phonon exchange, the low density edge of the dome is dictated by the vanishing of the density of states whereas the high density edge is dictated by the hardening of the phonon frequency (in conjunction with the Coulomb pseudopotential $\mu^*$).  
%would be suppressed by the low density of state on the underdoped side and by the effective interaction weakening due to the TO1 phonon stiffening on the overdoped side. For a further quantitative analysis, we need to obtain the electron-phonon vertex for the TO1 phonons. %(SBC: I am done with Section II for now...) 

\begin{figure}[h]
	\centering	
	\includegraphics[width=8.7cm]{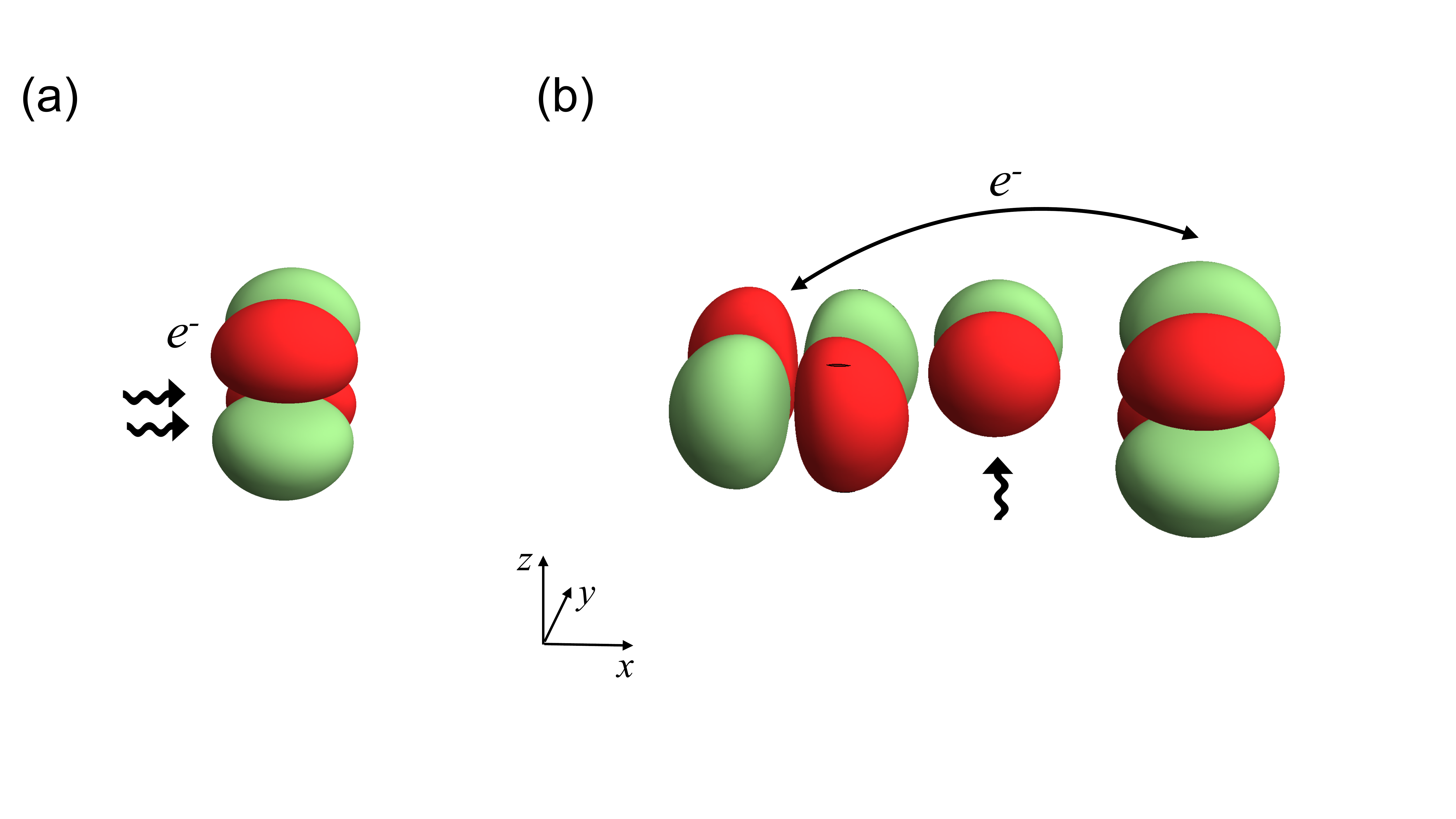}	
\caption{%Various 
	{\bf Electron coupling to the TO1 mode:} Two types of symmetry allowed couplings between $t_{2g}$ electrons and the TO phonons.  %We illustrate two orbitals displaced along the crystalline $x$ direction.  
	In a two phonon process (a), %electrons can tunnel between identical orbitals while absorbing or emitting
	a pair of TO phonons, each denoted by a wavy line oriented along the displacement axis, couples to the electron density. Single phonon processes (b) can occur provided the tunneling occur between two distinct $t_{2g}$ orbitals on the nearest neighbors.  Along the crystalline $x$ axis, a TO displacement along the y-direction mediates hopping between the $d_{xz}$ and $d_{yz}$ orbitals. %, wheras a TO displacement along the
	%$z$-direction couples the $d_{xy}$ and $d_{yz}$ orbitals.     
	The wavy line denotes the TO phonon displacement axis.}
	\label{orbitals}
\end{figure}

The only caveat in the hypothesis above is that it assumes a conventional coupling of the electrons to a single TO phonon.  Symmetry considerations however, require that if the initial and final electron states in a phonon exchange process have the {\it same} symmetry with respect to reflection about the plane normal to which the TO mode displacement occurs, the process must involve a pair of TO phonons\cite{Ngai1974} (for discussion on  superconductivity arising from such electron-phonon coupling, see Volkov {\it et al.}\cite{Volkov2021}). %While the possibility of superconductivity mediated by such coupling %providing the pairing mechanism for  %with the BCS ratio 
%has been discussed recently \cite{Volkov2021}, it gives us the BCS ratio %was shown to require the TO phonons to be 
%in the anti-adiabatic limit in contradiction with the SrTi$_{1-x}$Nb$_x$O$_3$ experimental results \cite{Bauerle1980,Yoon2021}. %rendering the scenario irrelevant to 
%(SBC: Would this be better as a footnote?) 
As we discuss below, the way around this constraint is to include multiple orbitals; a single TO phonon can scatter an electron from an orbital that is even under such a reflection to one that is odd, and {\it vice-versa}  \cite{Gastiasoro2021}.  
As we show, such processes can naturally account for a superconducting dome in this system.

\subsection{Pairing from TO phonon scattering} The qualitative effect of the electronic coupling to the single TO1 phonon can be most simply obtained from a microscopic model for the SrTiO$_3$ electronic band structure that incorporates the titanium (Ti) $3d$ $t_{2g}$ orbitals while assuming the simple %tetragonal 
cubic lattice structure. %{\bf I changed word from cubic to tetragonal. Fernandes' review also has a word ``pseudocubic" for it --YY}. 
The low-energy band structure can be described well by a minimal tight-binding model whose $k$-space representation can be written as \cite{Bistritzer2011,vanDerMarel2011,ZZhong2013,Gastiasoro2020A} 
\begin{align}
    H_0 =& \sum_{{\bf k},\alpha,s} (\epsilon_\alpha-\mu) c^\dagger_{{\bf k},\alpha,s} c_{{\bf k},\alpha,s}\nonumber\\ 
    &- \frac{\xi}{2}\sum_{{\bf k},\alpha,\beta,s,s'} {\bm \ell}_{\alpha\beta} \cdot {\bm \sigma}_{s,s'} c^\dagger_{{\bf k},\alpha,s} c_{{\bf k},\beta,s'},
    \label{3orbitalBand}
\end{align}
where $\alpha,\beta=X,Y,Z$ refer, respectively, to the Ti $d_{yz}$, $d_{xz}$, $d_{xy}$ orbitals, $s,s'$ the spin indices, $\xi$=19.3 meV from the Ti atomic spin-orbit coupling with the totally antisymmetric tensor $\ell^\alpha_{\beta\gamma} \equiv -i \varepsilon_{\alpha\beta\gamma}$ representing the effective $L=1$ orbital angular momentum of the TI $t_{2g}$ orbitals, and 
\begin{equation}
\epsilon_\alpha=\epsilon_0%+\Delta_\alpha
+4t_1\sum_{\beta\neq\alpha}\sin^2{\frac{k_\beta}{2}}+4t_2\sin^2{\frac{k_\alpha}{2}},
\end{equation}
is the intra-orbital hopping (with $\epsilon_0$ = 12.2 meV) whose $t_1>t_2$ anisotropy (0.615 eV and 0.035 eV, respectively) can be attributed to the quantum mechanical effect of the Ti $t_{2g}$ orbital symmetry \cite{vanDerMarel2011,Gastiasoro2020A}. %$\Delta_{X,Y,Z}=-2.2\times[1,1,-2]$meV is the tetragonal crystal field {\bf tetragonal crystal field is mentioned here. --YY}.

%, together with the fact that the hopping between the nearest Ti atoms occurs mostly through the intermediate O $p$-orbital 
The form of the electronic coupling to the TO1 phonons is determined by the interplay between the $t_{2g}$ orbital symmetry and the crystalline structure. As shown in Fig.~\ref{orbitals}, 
%1, 
the tunneling between different $t_{2g}$ orbitals between nearest neighbors is forbidden by inversion symmetry in a static lattice, but the TO1 mode displacements break inversion symmetry and thereby induce odd-parity inter-orbital tunneling.  %Given that the three TO1 modes displacement breaks 
%by changing the relative position of the Ti and O atoms in a TiO$_6$ octahedron, 
Given its odd-parity, this tunneling at the long-wavelength limit can be described by the following electron-phonon interaction \cite{SRPark2011,Volkov2020} %, which possesses cubic symmetry
	\begin{equation}
	H_{e-p}=g\sum_{{\bf k},{\bf q}}\sum_{i,\alpha,\beta,s}{\bm \phi}_{\bf q}\cdot[{\bm \ell}_{\alpha\beta}\times({\bf k}+{\bf q}/2)] c^\dagger_{{\bf k+q},\alpha,s}c_{{\bf k},\beta,s}
	\label{Eep}
	\end{equation}
where ${\bm \phi}$ is the TO1 mode displacement vector. The simplest justification for this coupling is to consider a uniform ${\bm \phi}\parallel {\bf \hat{z}}$, which %physically 
would %mean the uniform 
displace %of 
the Ti atom from the center of TiO$_6$ octahedron along the $z$-direction by a constant amount; this will turn on the nearest-neighbor hopping between the $d_{xy}$ and $d_{yz}$ ($d_{xz}$) in the $x$($y$)-direction through the O $p_y$ ($p_x$) orbital. The following two aspects of this electron-phonon coupling makes it viable as the pairing glue for superconductivity. 

First, the electron-phonon coupling of Eq.~\eqref{Eep} %in no sense is rendered irrelevant for being linear in gradient. %This is due both to this coupling remaining finite in the $q \to 0$ limit, hence allowing for the electronic coupling to the incipient ferroelectricity \cite{YWang2016}, and to the $k \to 0$ limit not involved the low energy physics given that superconductivity arises only at finite doping [THIS SENTENCE IS LONG AND CONFUSING - REVISE]. {\it here is my proposed revision to the sentence above--SR...  \\
is %therefore 
distinct from acoustic phonons 
coupling derivatively to the fermions. 
As long as there is a non-zero fermion density, the typical fermion momentum $\vert \bm k \vert \sim k_F$ is finite, and the electron-phonon coupling in Eq.~\eqref{Eep} survives even in the $q\rightarrow 0$ limit. %; the coupling used here  
%On the other hand, , %Eq.~\eqref{Eep} should not be taken as implying 
%However, %this does not imply  %effectively 
%growing %monotonically 
%with the %doping. %This can be justified as the superconducting dome has a finite lower critical doping and the screening of the electron-phonon coupling also increases with doping,
%increasing as fast as $n^{1/3}$ with the fermion density $n$, as the electron-phonon coupling is subject 
%due to 
%%the screening %at a 
%when $n$ is 
%at finite %carrier 
%fermion density %(see %an example is %computed 
%given in 
%App.~\ref{screening} for an example),  
%Altogether, we argue that we are justified in maintaining an %doping
%$n$-independent electron-phonon coupling strength as in Sec. II in our calculation of the superconducting dome.
%would be qualitatively better model by the constant 
%we expect the 
%effective electron-phonon coupling $gk_F$ strength %to be 
%essentially independent of the density 
%rather than the constant $g$.
In the interest of simplicity, we consider the case where $gk_F$ is independent of density; what this implies will be discussed upon obtaining the effective BCS interaction.

Second, the electron-phonon coupling of Eq.~\eqref{Eep} %leads to %induces a Rashba-type 
can induce an intra-band pairing interaction 
%coupling to the TO1 phonon  %would arise from the combination of the electron-phonon coupling of  and 
%due 
due to the presence of atomic spin-orbit coupling in Eq.~\eqref{3orbitalBand} \cite{Brydon2014,Scheurer2016,MLee2020}. This is %most straightforward in 
not limited to the three C$_3$ rotational axes of the cubic lattice, %{\bf "pseudo-cubic" might be more precise, since we are talking about low temperature. --YY} %(111), ($\bar{1}$11), (1$\bar{1}$1), ($\bar{1}\bar{1}$1) directions 
where the eigenstates of the $H_0$ of Eq.~\eqref{3orbitalBand}  are the effective $j=1/2$ and $j=3/2$ states (The projections of ${\bm \phi}\cdot ({\bm \ell} \times {\bf k})$ to the $j=1/2$ and the $j=3/2$ subspaces %would be 
are $\frac{4}{3}{\bm \phi}\cdot({\bf j} \times {\bf k})$ and $-\frac{2}{3}{\bm \phi}\cdot({\bf j} \times {\bf k})$, respectively \cite{HKim2018,Venderbos2018}.) Even away from this band degeneracy, the intra-band Rashba coupling of the TO1 phonon \cite{Kozii2015,YWang2016,Gastiasoro2020B} can be obtained by treating the sum of Eq.~\eqref{Eep} and the atomic spin-orbit coupling of Eq.~\eqref{3orbitalBand} as  perturbations \cite{MKim2016,MLee2020}.

\begin{figure}[h]
	\centering	\includegraphics[width=8.52cm]{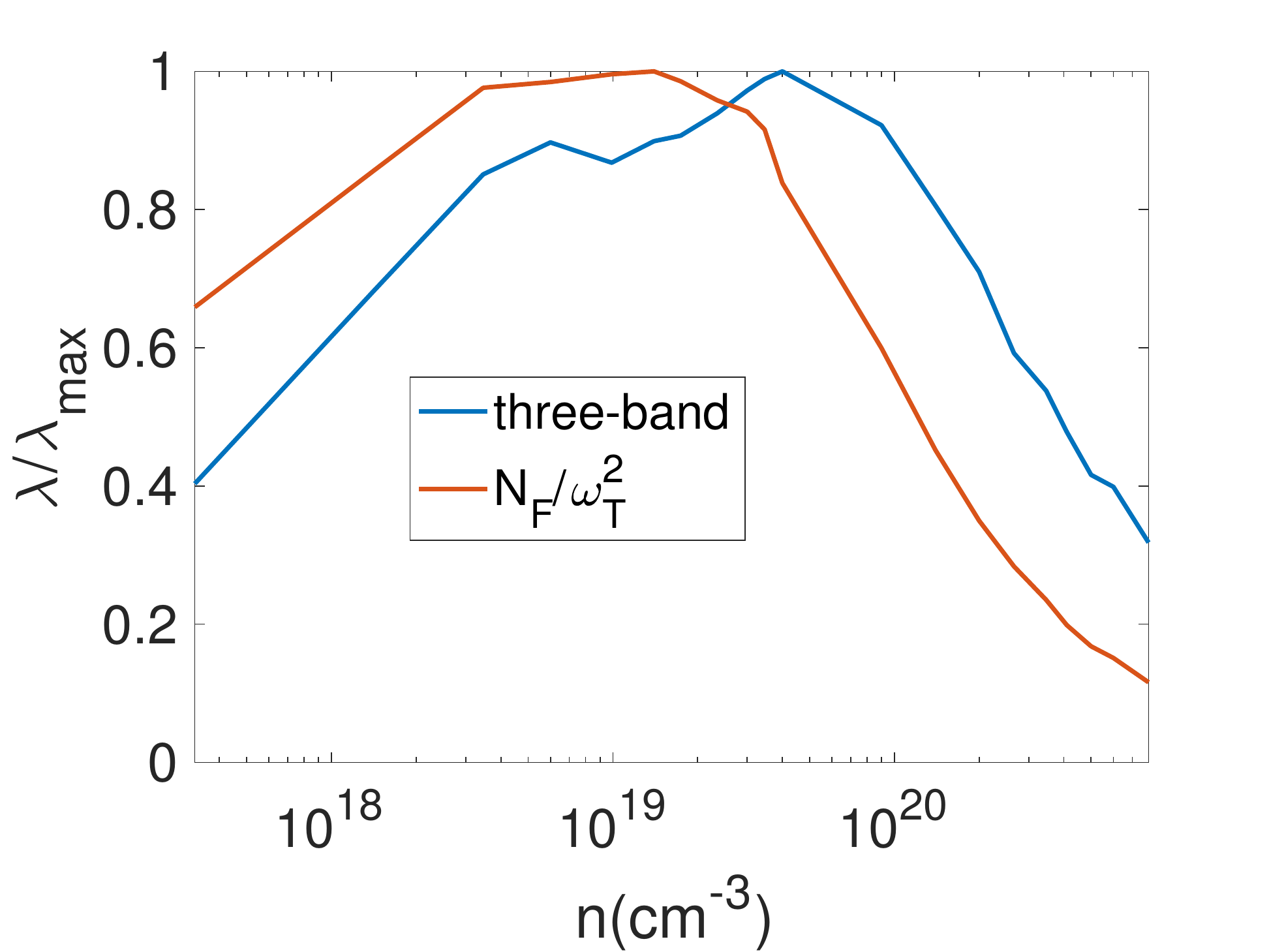}	
	\caption{{\bf BCS eigenvalue} ${\bm \lambda_{\rm {\bf BCS}}}${\bf :}  Comparison between the BCS eigenvalue $\lambda_{\rm BCS}$ dome in the three-orbital model of Eq.~\eqref{3orbitalBand} and the single-band model of Eq.~\eqref{singleBand}. In the former, the weak tetragonal crystal field strength of $\sim$2.2 meV \cite{Gastiasoro2020A} has been included, which has little effect around the optimal doping. For a better comparison on the superconducting dome, both lines are rescaled with respect to their maximal value, {\it i.e.} $\lambda/\lambda_{\rm max}$ is the rescaled BCS eigenvalue.  Red line is the result from single-band estimation, while blue line is the result from three-band calculation.}
	\label{f1}
\end{figure}

We now derive the dimensionless effective BCS pairing interaction from this electron-phonon coupling using the %following 
Dyson's equation in the Nambu space where the electron self-energy arises entirely from the Cooper pairing. For this Dyson's equation, 
\begin{align}
    \Sigma({\bf k}, i\nu_m) =& 2\frac{k_B T}{\hbar}\sum_{{\bf k}',i,j} \chi_{ij}({\bf k}-{\bf k}')\nonumber\\
    &\times F_j({\bf k},{\bf k}'-{\bf k})\mathcal{G}({\bf k}',i\nu_m) F_i({\bf k},{\bf k}'-{\bf k}),
    \label{Dyson}
\end{align}
%for the electron self-energy, 
where $\nu_m \equiv (2m+1)\pi k_B T/\hbar$ (with $m \in \mathbb{Z}$) is the fermionic Matsubara frequency, $\mathcal{G}$ the electronic Green's function and
\begin{equation}
	F_i({\bf k},{\bf k'-k})\equiv{g}[{\bm \ell}\times({\bf k}+{\bf k'})]_i/4
%\label{e5}
\end{equation}
is the electron-phonon interaction vertex from Eq.~\eqref{Eep}, with $g\propto{n^{-1/3}}$ to maintain an %n-independent 
effective electron-phonon coupling strength independent of the carrier density $n$. We take advantage of the adiabaticity of the TO1 phonon to ignore the boson dynamics \cite{Gastiasoro2020B} and take the static TO1 propagator,
\begin{equation}
    \chi^{ij}({\bf q}) =\langle \phi^i_{-{\bf q}} \phi^j_{\bf q} \rangle=  \frac{\hbar}{M_T}\frac{\delta_{ij}-\hat{q}_i. \hat{q}_j}{\omega^2_T + c_T^2 q^2},
\end{equation}
%the TO1 propagator. %(Note: overall factor of $c_T$ is removed, since eventually we take it to be zero). 
%hence confine ourselves to the frequency independent pairing. 
where $M_T$ is the TO1 phonon effective mass. Given that the self-energy for this Dyson's equation is given as the linear combination of the pairing gap, we need to consider the form of pairing gap that would be favored by the electron-phonon coupling of Eq.~\eqref{Eep}. %With regards to the orbital dependence, we note that the orbital hybridization occurs only in small portions of the Fermi surface, while 
With regards to the Cooper pair spin states, we note that any electron-phonon coupling, even with odd-parity, favors spin-singlet pairing \cite{Frigeri2004,Brydon2014,YWang2016,Scheurer2016}. Therefore our pairing gap should be intra-band, even-parity, pseudospin-singlet (frequency-independence being already assumed by Eq.~\eqref{Dyson}), giving us
%\begin{equation}
%\Sigma_n({\bf k})=\tau^x\sum_{\alpha}\Delta_\alpha({\bf k})(i\sigma^y\delta_\alpha),
%\label{gap}
%\end{equation}
\begin{equation}
\Sigma({\bf k})=\tau^xu({\bf k})\Delta({\bf k})(i\sigma^y\delta[\alpha_{\bf k}])u^T({\bf -k}),
\label{gap}
\end{equation}
written in orbital basis. Here, $\delta[\alpha_{\bf k}]$ is a $3\times3$ matrix in  band space, with unity at the $(\alpha_{\bf k},\alpha_{\bf k})$ element and zero elsewhere for state {\bf k} on band $\alpha_{\bf k}$, and $u({\bf k})$ is the unitary transformation that diagonalize the normal state Hamiltonian. Hence by taking the one-loop approximation to the electronic Green's function
\begin{equation}
    \mathcal{G}({\bf k}, i\nu_m) = \mathcal{G}_0({\bf k}, i\nu_m) + \mathcal{G}_0({\bf k}, i\nu_m) \Sigma({\bf k}, i\nu_m) \mathcal{G}_0({\bf k}, i\nu_m),
\end{equation}
where the $\mathcal{G}^{-1}_0({\bf k}, i\nu_m)=\frac{1}{2}(i\nu_m - \tau^z h_{\bf k})$ is the bare electron Green's function (with $h_{\bf k}$ being the $3\times3$ tight-binding Hamiltonian of Eq.~\ref{3orbitalBand} in the orbital basis), the Dyson's equation of Eq.~\eqref{Dyson} can be readily reduced to the linearized gap equation of the form %$\Delta_\alpha ({\bf k}) = \sum_{{\bf k}',\beta} V_{\alpha\beta} ({\bf k},{\bf k}') \Delta_\beta ({\bf k}')$
$\Delta ({\bf k}) = \sum_{{\bf k}'} V ({\bf k},{\bf k}') \Delta ({\bf k}')$
whose eigenvalues represent the dimensionless effective BCS pairing interactions for the pairing channels satisfying Eq.~\eqref{gap}; details are given in Methods. %App. \ref{linearizedGap}.

The effective BCS interaction of the above three-band model derived from this procedure %can be found 
plotted in Fig.~\ref{f1} with comparison with the %Sec. II 
single-band estimation of Eq.~\eqref{singleBand}, demonstrates that the superconducting dome arises also with the unconventional, {\it i.e.} odd-parity, electron-phonon coupling of Eq.~\eqref{Eep}. While the optimal doping (and therefore chemical potential) value may be shifted, the suppression of superconductivity on the low density dome edge by the vanishing density of states and on the high density dome edge by the TO1 phonon hardening still remains. This remains qualitatively true as long as there is any nonzero screening effect on Eq.~\eqref{Eep} %is accounted for, 
%{\it i.e.} %the effective electron-phonon coupling strength
%$gk_F$ does not increase as fast as $n^{1/3}$  
%$g$ decreasing 
%with the %dopant concentration 
%carrier 
%density $n$, %one 
%the attenuation of 
that attenuates $g$ at sufficiently high density  (For $\langle{\bm \phi}_{{\bf q}=0}\rangle \neq 0$, Eq.~\ref{Eep} gives us the inversion symmetry breaking electron hopping, %which has been shown to increase with the internal electric field seen at the Ti-O bond \cite{Shanavas2016}. 
one of the key ingredients for the Rashba effect \cite{SRPark2011, MKim2016}. %A finite carrier concentration will attenuate $g$ as it will screen the internal electric field for a given $\langle{\bm \phi}_{{\bf q}=0}\rangle$. 
The reduction of the Rashba coefficient with the increasing carrier concentration found in the recent first-principle calculation for Bi$_2$WO$_6$, a related material \cite{Djani2019}, and the experiment on the few-layer GeTe\cite{XYang2021} are possible instances of screening attenuating such electron hopping and by extension, the parameter $g$ in Eq.~\ref{Eep}.), 
the simplest modeling of which is the density-independent $gk_F$ we have used for obtaining Fig.~\ref{f1}.

\subsection{Normal state considerations}
We shall now show that the above phonon-mediated superconducting mechanism is consistent with $T^2$ resistivity that has been observed for SrTiO$_3$ at low doping \cite{Lin2015}. %[YUE, IN THIS SECTION YOU SHOULD ALSO CITE THE FEW PAPERS I SENT YOU ON THIS TOPIC, perhaps also Maslov, AND PERHAPS STATE THOSE WHICH ARE SIMILAR/DISTINCT IN SCENARIO.] 
This behavior has attracted attention because, given the small Fermi surface, it cannot be sufficiently explained by the electron-electron scattering process  \cite{Kumar2021, verma2014, zhou2018}. Our point here is that this behavior can be actually explained from the single-phonon electron-phonon scattering process, the imaginary part of the self-energy is given by the Fermi's golden rule (see %App.~\ref{self-energy} 
Sec.~\ref{self-energy} for derivation):
\begin{equation}
\begin{split}
    \text{Im}\Sigma({\bf k},\epsilon)\!=\!&-\!\pi\!\sum_{\bf q}| g_{{\bf k'},{\bf k}}|^2\left[\delta(\epsilon\!-\!\xi_{\bf k'}\!+\!\omega_{\bf q})(n_F(\xi_{\bf k'})\!+\!n_B(\omega_{\bf q}))\right.\\
    &+\left.
    \delta(\epsilon\!-\!\xi_{\bf k'}\!-\!\omega_{\bf q})(n_F(-\xi_{\bf k'})\!+\!n_B(\omega_{\bf q}))
    \right],
    \end{split}
    \label{ImSelf}
\end{equation}
where $\xi_{\bf k}$ and 
$\omega_{\bf q}$ are 
the dispersions of electrons and 
phonons. $g_{{\bf k'},{\bf k}}$ is the electron-phonon coupling strength (with ${\bf k'} = {\bf k}+{\bf q}$). $n_F$ and $n_B$ are the Fermi and Bose distributions. %Optical phonons are found to be weakly dispersed: 
For the optical phonons, the Einstein model is sufficient to capture the qualitative behavior, {\it i.e.} $\omega_{\bf q}\approx\omega$. Therefore, if we focus on electrons at the Fermi surface, the self-energy %simplifies to
from scattering off a single branch of optical phonons would be
\begin{equation}
\text{Im}\Sigma({\bf k},\epsilon\!=\!0)\!=\!-2\pi\left[(n_F(\omega)\!+\!n_B(\omega))\right]\sum_{\bf q}|g_{{\bf k'},{\bf k}}|^2\delta(\xi_{\bf k'}\!-\!\omega),
\end{equation}
from which the relation between the scattering rate and the self-energy $1/\tau=-2{\rm Im}\Sigma/\hbar$ gives us the scattering rate formula of the form
\begin{equation}
    1/\tau=A\left[(n_F(\omega)+n_B(\omega))\right].
    \label{scatteringSingleBranch}
\end{equation}
%for a single branch of optical phonons with frequency $\omega$. 
The simplified scattering rate depends on the phonon energy $\omega$, temperature $T$ and a coefficient A, proportional to the magnitude square of the electron-phonon coupling strength. By itself, Eq.~\eqref{scatteringSingleBranch} cannot give rise to the $T^2$ resistivity; the resistivity rather shows a $T$-linear behavior at high temperature $T\gg\omega$ %as well as 
and an exponential suppression at low temperature $T\ll\omega$ with a crossover regime for $T \sim \hbar\omega/k_B$. 

\begin{figure}[h]
	\centering	\includegraphics[width=8.52cm]{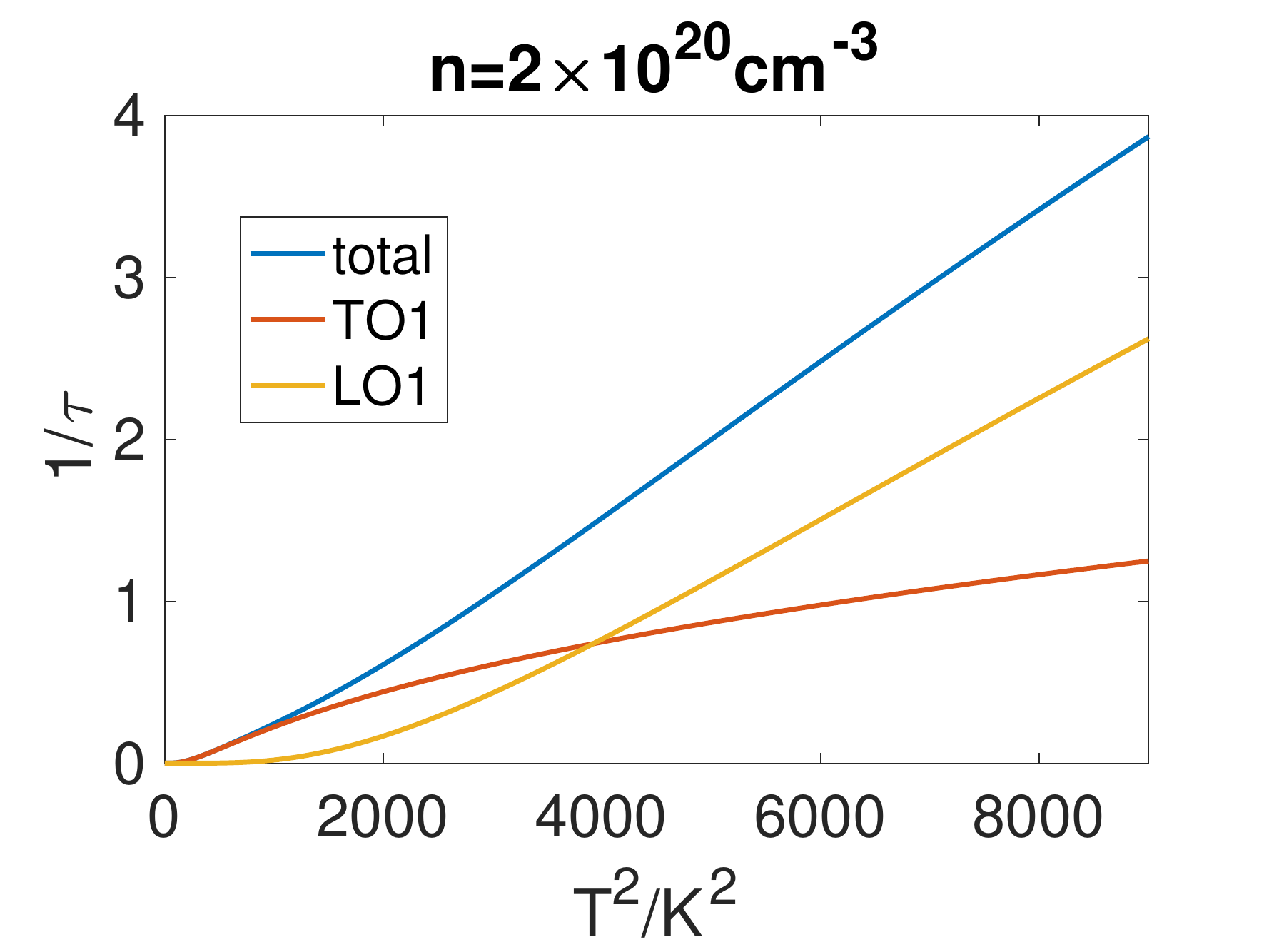}	\caption{{\bf ${\bf T^2}$ scattering rate:} (Blue) Scattering rate as a function of $T^2$. Contributions from TO1 (Red) and LO1 (yellow) phonons are included. Parameters $\omega_T=6$meV, $\omega_L=20$meV \cite{Yoon2021}, $A_T=1$ and  $A_L=15$ are taken. Blue line is the sum of the red (TO1 contribution) and
	yellow (LO1 contribution) line.}
	\label{f2}
\end{figure}

The $T^2$ resistivity can nevertheless arise from scattering by multiple branches of optical phonons at different frequencies. As charge carrier density increases, the energy of the TO1 phonon increases from 20K to 100K in the $T^2$ resistivity measurement for Nb-doped STO \cite{Lin2015}. Energies of other optical phonons are essentially doping-independent. Among them, the LO1 phonon has the lowest energy at approximately 200K. Due to the strong electron-phonon coupling in the LO1 phonon channel \cite{Swartz2018}, its contribution to the scattering rate should not be neglected. We thus have the combined electron-phonon scattering rate:
\begin{equation}
\begin{split}
    1/\tau=A_T\left[n_F(\omega_T,T)+n_B(\omega_T,T)\right]\\+A_L\left[n_F(\omega_L,T)+n_B(\omega_L,T)\right]
\end{split}\label{Eqscatter}
\end{equation}
with $A_L\gg{A}_T$. This leads to a broad crossover regime starting from TO1 phonon energy, up to LO1 phonon energy. This crossover regime could give an approximate $T^2$ scattering rate, as shown in Fig.~\ref{f2} for a doping at the superconducting dome. The temperature is much lower than the Fermi energy at this doping. %Realistic phonon energies are considered \cite{Yoon2021}. 
At higher temperature, including electrons away from the Fermi surface may be needed for the computation of the scattering rate. Approximate $T^2$ resistivity at other dopant concentrations can be found in Appendix~\ref{scattering}. 
To summarize, %this section, 
we note that while other mechanisms are possible \cite{Kumar2021}, %[cite Maslov], 
the measured phonon frequency values along with the calculations presented here make it impossible to rule out a scenario in which the coupling to both TO1 and LO1 phonon modes can produce the $T^2$ resistivity at least over a range of temperatures (Fig.~\ref{f2}).

\section{Discussion} %discuss other quantum paraelectrics here, talk about 2d vs 3d.  
%Our present discussion %on the superconductivity of the doped paraelectric SrTiO$_3$ 
%should provide %us with 
%a starting point for discussing superconductivity in other quantum paraelectrics. In particular, superconductivity is reported in n-doped KTaO$_3$ only at the interface and not in the bulk \cite{Ueno2011,ZChen2021L,CJLiu2021,ZChen2021S} despite Ta possessing much stronger spin-orbit coupling than Ti.

In this paper, we have utilized  experimental data to constrain and deduce the most likely pairing mechanism underlying Nb-doped SrTiO$_3$.  Such strategies can perhaps be of broader relevance to other materials such as PbTe and KTaO$_3$ that are close to a ferroelectric transition.  It would be of considerable interest to repeat such planar tunneling measurements in these systems.  For example, the ideas presented here can help shed light on the recent observations of interfacial superconductivity in  KTaO$_3$, which shows a surprisingly high $T_c$ of $\sim$2  K, while showing no signs of bulk superconductivity at the present time.  At an interface, the presence of Rashba spin-orbit coupling allows for the %conventionally
%linear 
coupling to a single TO1 phonon as in Eq.~\eqref{Eep}. The effective strength of the phonon coupling is enhanced by bulk spin-orbit coupling, which may account for the enhancement of superconductivity at the interface of this system.  Further experimental studies in similar materials may help uncover the global phase diagram of quantum paraelectrics as a function of spin-orbit strength, dopant concentration, and spatial dimensionality. 

\section{Methods}
%\label{linearizedGap}
\subsection{BCS Eigenvalue calculation}

%By performing matrix inner product with the basis matrix $M_\alpha$ on both side, we reach 
%This Ansatz allows us to derive 
%In deriving the 
The linearized gap equation 
%\begin{equation}
%\Delta_\alpha({\bf k})=\sum_{{\bf k'}\beta}V_{\alpha\beta}({\bf k,k'})\Delta_\beta({\bf k'}),
%\end{equation} 
\begin{equation}
\Delta({\bf k})=\sum_{{\bf k'}}V({\bf k,k'})\Delta({\bf k'}),
\end{equation}  
from the Dyson's equation of Eq.~\eqref{Dyson} need to have %where
\begin{widetext}
%$$
	%V_{\alpha\beta}({\bf k,k'})=CN_F\sum_{ij\mu}\chi_{ij,0}({\bf k'-k})\text{Tr}\{M^\dagger_\alpha F_j({\bf k},{\bf k'-k})u({\bf k'})[u^\dagger({\bf k'})M_\beta{u}^*({-\bf k'})]_{\mu\mu}u^T({-\bf k'})F_i({\bf k},{\bf k'-k})\},
%$$
\begin{equation}
V({\bf k,k'})=CN_F\sum_{ij}\chi_{ij,0}({\bf k'-k})\text{Tr}\{M^\dagger({\bf k}) u^\dagger({\bf k}) F_j({\bf k},{\bf k'-k})u({\bf k'})M({\bf k'}) u^T({-\bf k'})F_i({\bf k},{\bf k'-k})u^*({\bf -k})\},
\end{equation}
\end{widetext}
where $M({\bf k}) \equiv i\sigma^y\delta[\alpha_{\bf k}]$, while the constant $C$ is a function of energy cutoff and critical temperature, {\it i.e.} $C\propto\log(\omega_c/T_c)$. %Note that we are keeping only the intra-band pairing terms from band $\mu$, by applying projection operator $[...]_{\mu\mu}$ onto band $\mu$.
The eigenvalues of the linearized gap equation are exactly $\lambda_{\rm BCS}$'s that determine $T_c$. 
%Numerically, the above linearized gap equation can be treated as an eigenvalue equation of matrix $V_{\alpha\beta}({\bf k,k'})$, in a vector space spanned by N momenta with 3 orbits.} 
Numerically, the above linearized gap equation can be treated as an eigenvalue equation of matrix $V({\bf k,k'})$, in a vector space spanned by N momenta. $T_c$ is determined by the largest eigenvalue, and the corresponding eigenvector (dominant pairing channel) is s-wave. The Fermi energy $E_F$, the carrier density $n$ and the TO1 phonon energy $\omega_T$ are taken from the tunneling experiment \cite{Yoon2021}. We assume $N_F,k_F\propto\sqrt{E_F}$, and $c_T=0$.

\subsection{Scattering rate derivation}
\label{self-energy}

The formula for the scattering rate of electrons due to a single branch of phonons, as given in Eq.~\eqref{scatteringSingleBranch}, 
can be derived from the one-loop electron self-energy. The generalized form of the electron-phonon coupling,
\begin{equation}
H_{\rm e-ph} = \sum_{{\bf q},\lambda}\sum_{\bf k} g_{\lambda;ss'} ({\bf q},{\bf k})(a_{{\bf q},\lambda}+a^\dagger_{-{\bf q},\lambda})c^\dagger_{{\bf k}+{\bf q},s} c_{{\bf k},s'},
\end{equation}
where $a (a^\dagger)$ is the phonon annihilation (creation) operator and $\lambda$ the phonon polarization, can be taken as the starting point %from which one can use the bare electron and phonon propagator
%\begin{align*}
%    G_0 ({\bf k},\omega) =& \frac{1}{\omega - (\epsilon_{\bf k}-\mu)\hbar},\nonumber\\
%    D_0 (\lambda;{\bf q},\nu) =& \frac{2\omega_{{\bf q},\lambda}}{\nu^2 - \omega_{\lambda;{\bf q}}^2},
%\end{align*}
to obtain the one-loop electron self-energy
\begin{widetext}
\begin{align}
  \Sigma ({\bf k},i\omega_n) =&\frac{k_B T}{\hbar}  \frac{1}{\hbar}\sum_{\bf q}|g_\lambda ({\bf q},{\bf k})|^2 \sum_{i\nu_m} \frac{1}{i\omega_n + i\nu_m - \xi_{{\bf k}+{\bf q}}} \frac{2\omega_{{\bf q},\lambda}}{\nu_m^2 + \omega_{\lambda;{\bf q}}^2}\nonumber\\  
  =& \frac{1}{\hbar}\sum_{\bf q}|\tilde{g}_\lambda ({\bf q},{\bf k})|^2 \left[\frac{n_B (\omega_{\lambda;{\bf q}},T)+n_F (\xi_{{\bf k}+{\bf q}},T)}{i\omega_n +\omega_{\lambda;{\bf q}} -\xi_{{\bf k}+{\bf q}}}+\frac{n_B (\omega_{\lambda;{\bf q}},T)+n_F (-\xi_{{\bf k}+{\bf q}},T)}{i\omega_n -\omega_{\lambda;{\bf q}} -\xi_{{\bf k}+{\bf q}}}\right],
\label{self}
\end{align}
\end{widetext}
where $\omega_{{\bf q},\lambda}$ is the phonon eigenfrequency, in the Matsubara frequency. One can see see how Eq.~\eqref{ImSelf} 
%of the main text 
can be obtained by taking the imaginary part of Eq.~\eqref{self}. Applying the Einstein model for phonons with $\omega_{\bf q} = \omega_0$ for all {\bf q}, the scattering rate of Eq.~%(6) 
\eqref{scatteringSingleBranch} 
%of the main text 
is obtained the imaginary part of the electron self-energy
\begin{equation}
\frac{1}{\tau} = -\frac{2}{\hbar}{\rm Im} \Sigma(\omega+i\delta \to 0),
\end{equation}
with
\begin{equation}
A = \frac{2\pi}{\hbar^2} \sum_{\bf k'}|\tilde{g} ({\bf k}'-{\bf k},{\bf k})|^2[\delta(\xi_{{\bf k}'}/\hbar-\omega_0)+\delta(\xi_{{\bf k}'}/\hbar+\omega_0)].
\end{equation}

\section{Data availability} 
Relevant data in this paper are available upon reasonable request.

\section{Acknowledgement} 
We thank Piers Coleman, Rafael Fernandes, Changyoung Kim, Dmitrii Maslov, and Hyeok Yoon for useful discussions. %S.R was supported by the Department of Energy, Office of Basic Energy Sciences, Division of Materials Sciences and Engineering, under contract No. DEAC02-76SF00515. 
Y.Y., H.Y. H., and S.R. were supported by the Department of Energy, Office of Basic Energy Sciences, Division of Materials Sciences and Engineering, under contract No. DEAC02-76SF00515. 
S.B.C. was supported by the National Research Foundation of Korea (NRF) grants funded by the Korea government (MSIT) (2020R1A2C1007554) and the Ministry of Education (2018R1A6A1A06024977).

{\it Notes added} - While drafting this manuscript, we have learnt of the recent preprint by Gastiasoro {\it et al.} \cite{Gastiasoro2021} which %also considers  essentially the same microscopic mechanism for superconductivity in the doped SrTiO$_3$ considered here but attempts to show how this mechanism can give the experimentally observed $T_c$, a question not considered here, but does not deal with the main question addressed here, namely the evolution of superconductivity with doping.
which has some overlap with the ideas presented here. However, our motivation here is distinct in the use of recent tunneling experiments to constrain pairing mechanisms.

\section{Author information}

{\it Contributions:} Y. Y. performed numerical calculation. All authors contributed to designing the project and writing the manuscript.

{\it Corresponding authors:} Correspondence to S. Raghu (sraghu@stanford.edu) or Suk Bum Chung (sbchung0@uos.ac.kr)

%\section{Ethics declarations}
\section{Competing interests}

%{\it Competing interests:} 
The authors declare no competing interests.

%\bibliography{sto}
%\bibliographystyle{naturemag}

\newpage

\appendix

\section{Supplementary Figures}
\label{scattering}

We present the doping-dependence of scattering rate under Eq.~(7) of the main text. %\ref{Eqscatter}. 
Approximate $T^2$ resistivity is found for various dopant concentrations. 

\begin{figure}[h]
	\centering	\includegraphics[width=8.52cm]{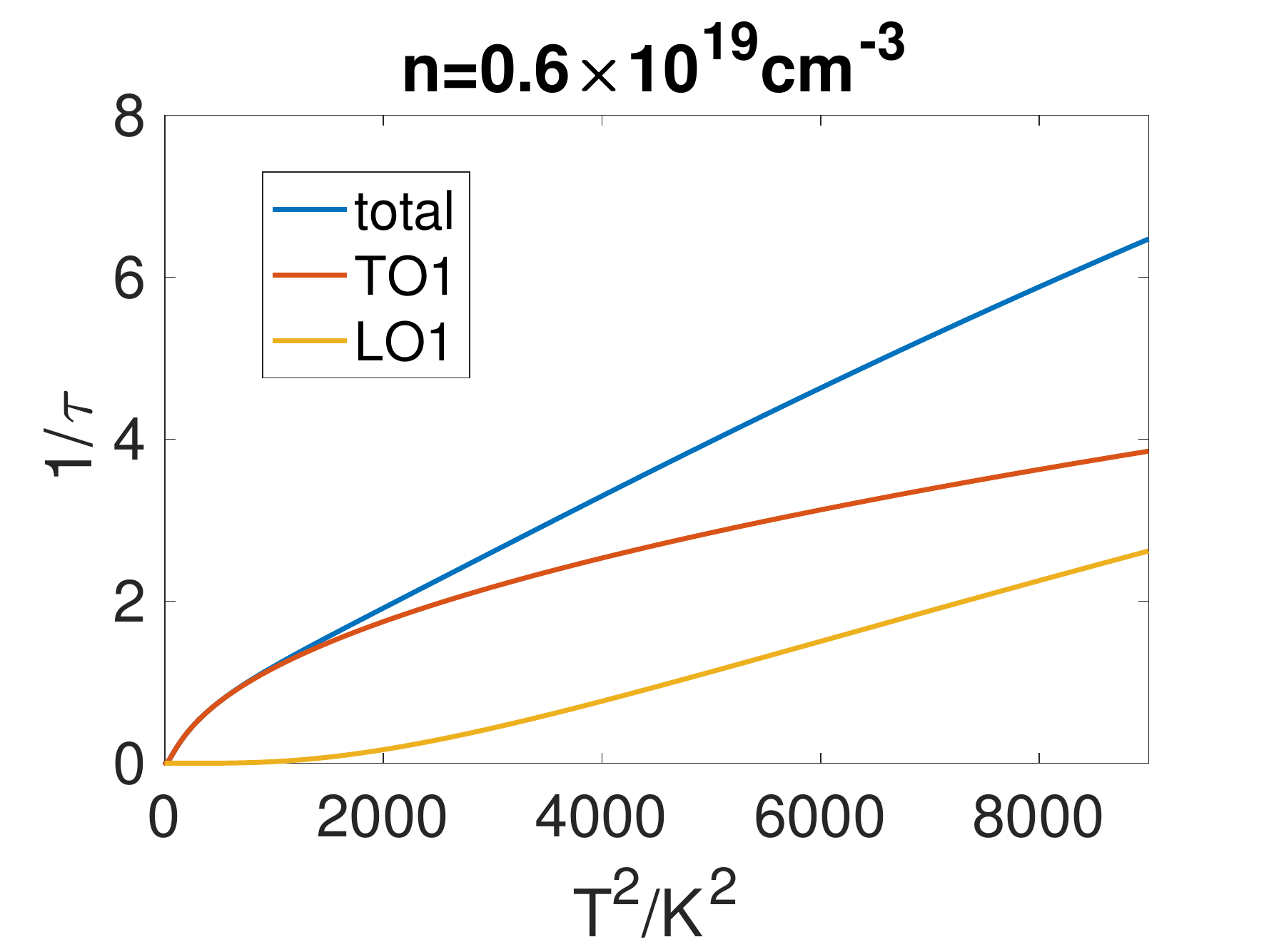}
	\centering \includegraphics[width=8.52cm]{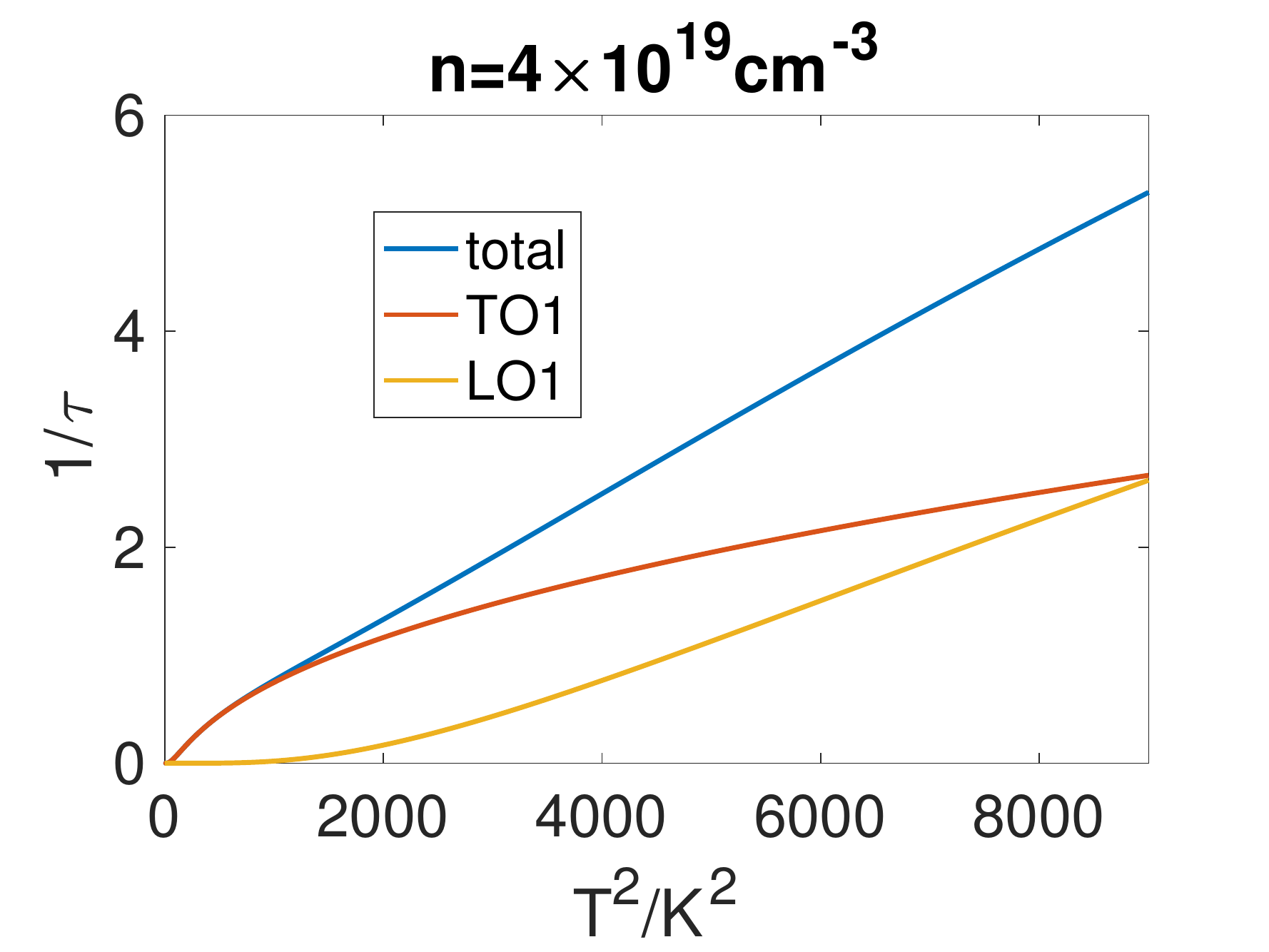}
	\caption{Scattering rate at different dopings. $\omega_T=2.1$meV (left) and $\omega_T=3$meV (right) are taken respectively. %for the figure on the left and the right. 
	Other parameters are the same as in main text. }
\end{figure}

\end{document}